\documentstyle[11pt]{article}
\begin{document}
\title{
        Magnetization Process of One-Dimensional
         Quantum Antiferromagnet:
        The Product Wavefunction
         Renormalization Group Approach}
\author{Yasuhiro Hieida, Kouichi Okunishi, and Yasuhiro Akutsu}

\date{January 31, 1997}
\maketitle
\begin{center}
\it Department of Physics,
Graduate School of Science, Osaka University,\\
Machikaneyama-cho 1-1, Toyonaka, Osaka 560, Japan\\
\end{center}

{\bf Abstract}

The product-wavefunction renormalization group
 method, which is a novel numerical renormalization group scheme
proposed recently,
 is applied to one-dimensional quantum spin chains in a
 magnetic field.
  We draw the zero-temperature magnetization curve
 of the
 spin chains,
 which excellently agrees with
 the exact solution in the whole range of the field.
\begin{itemize}
\item[\bf PACS codes:]  02.70.-c, 05.30.-d, 75.10.Jm, 75.30.Kz
\item[\bf Key Words:] quantum spin chain, numerical renormalization,
magnetization curve
\item[\bf Address:] Yasuhiro Hieida c/o Yasuhiro Akutsu
\item[]  Department of Physics, Graduate School of Science,
\item[] Osaka University, Toyonaka, Osaka 560, JAPAN
\item[\bf Phone:] +81-6-850-5349, {\bf Fax:} +81-6-845-0518
\item[\bf e-mail:] hieida@godzilla.phys.sci.osaka-u.ac.jp
\end{itemize}
\newpage

Many interesting problems which have been left unsolved mostly concerned
 with strongly correlated systems where many-body effects play essential
 roles. For these problems conventional many-body-theoretic techniques, e.g.,
 the mean-field approximation and the perturbation expansion,
 lose their power;
 these methods often fail in explaining even qualitative features of the
 system.  As
promising substitutes, we have direct numerical approaches
 which are rapidly developing in accordance with the remarkable advancement
 of the computer technology. Examples are the Monte-Carlo simulation, the
 molecular dynamics and the exact diagonalization of Hamiltonian
matrices
 or transfer
 matrices, which have been successfully applied to various problems.

In these direct numerical methods we face another problem: Since the system
 size $N$ (number of lattice points, for example) which we can deal with
 is always finite, we need extrapolation with respect to $N$ to extract
 the behavior in the thermodynamic limit ($N\rightarrow \infty$).  As a
 standard prescription we have the finite-size scaling method~\cite{FSS},
 which has been one of
 most successful extrapolation schemes for studying the
 critical phenomena. In applying the finite-size scaling, we must compute
 the $N$-dependence of physical quantities for large enough $N$, which often
 requires long computation time. For the exact diagonalization method, the
 large-$N$ problem is rather serious due to the exponential growth of the
 Hamiltonian matrix dimension, which imposes severe limitation on  feasible
 system size $N$.

The density-matrix renormalization group (DMRG) method invented by S. R.
 White is a very important step to overcome the large-$N$ difficulty. The
 method was introduced as an approximate diagonalization scheme and has
 been applied to one-dimensional quantum systems, with results demonstrating
 surprising efficiency of the method~\cite{orig-DMRG}.
 The method is also applicable to two-dimensional lattice statistical models
for diagonalization
 the transfer matrices instead of
 the Hamiltonian matrices~\cite{classical-DMRG}.

Quite recently, two novel algorithms are devised by T. Nishino and K.
 Okunishi, which are closely connected to the DMRG.
  One is the corner-transfer-matrix renormalization group (CTMRG)
 method~\cite{CTMRG,NOK}
and the other is the product-wave-function
 renormalization group (PWFRG) method~\cite{PWFRG}.
 In the CTMRG, a
 systematic renormalization scheme for  large-lattice-size two-dimensional
 classical spin systems is given in terms of Baxter's corner transfer
 matrix~\cite{Baxter}.
 In the PWFRG, one diagonalizes the ``wavefunction matrix'' instead of
 the density matrix to obtain the
``projection matrix'' for retained block-state
bases and
 sets up a recursion relation for the
``projection matrix''. As has been shown in
refs~\cite{CTMRG,NOK,PWFRG}.
 both methods are highly efficient for 2D classical systems.  Hence,
 if we can extend them to 1D quantum systems, similar efficiency may be
expected.  For 1D quantum systems, the PWFRG seems to be more easily
 implemented than the CTMRG, because the latter takes full advantage of
the two-dimensionality of the system.  The actual implementation, however,
 is non-trivial because the transfer-matrix multiplication employed in
 the original PWFRG lose meaning for quantum systems.

The aim of this article is to show that the PWFRG can actually be extended to
quantum systems and to demonstrate the efficiency of the ``quantum PWFRG'' by
applying it to 1D quantum antiferromagnets in
the uniform
 magnetic field to obtain the magnetization curve at zero temperature.
 The importance of the study of the magnetization curve itself is clear. First,
the magnetization process
 or $M$(magnetization)-$H$(magnetic field) curve is a directly and accurately
 observable quantity by experiments~\cite{experiments1,experiments2}.
 We need reliable theoretical calculation
 on various models for comparison with experiments. Second, a magnetized
 state which is the lowest energy state for $H>H_{\rm c}$ ($H_{\rm c}$:
 lower critical field ), is an excited state at $H=0$. This means that the
 magnetization curve contains a considerable
 amount of information about the whole energy-level structure of the system.
 Third, for gapless systems, the efficiency (or inefficiency) of
 the numerical
 renormalization groups like the DMRG and the PWFRG has not been fully
 tested yet.
  Calculations of the magnetization curve would also serve as this
test, because
 the finitely-magnetized ground states of Heisenberg-like quantum
 antiferromagnets are gapless states.

Consider a spin-$S$ antiferromagnetic spin chain in a field whose Hamiltonian
 ${\cal H}$ is expressed as
\begin{equation}
{\cal H}=\sum_{i}h(\vec{s}_{i},\vec{s}_{i+1})-H\sum_{i}s^{z}_{i}\label{eq:1}
\end{equation}
where $\vec{s}_{i}$ is the spin operator at the site $i$ and $h$ is the
nearest-neighbor
 coupling function (or {\em local Hamiltonian}).  We take a unit where
$g\mu_{\rm
 B}=1$ ($g$: $g$-factor, $\mu_{B}$: Bohr magneton) or these factors are
 absorbed into the field $H$.  The ``pure'' Heisenberg model corresponds
 to
$h=h(\vec{s}_i,\vec{s}_{i+1})=-J\vec{s}_i\cdot\vec{s}_{i+1}$
($J$: an exchange coupling
 constant), but we consider more general form of $h$ conserving the
total $S^{z}$.
 Our problem is to find the lowest energy state of ${\cal H}$.

Let us briefly explain the PWFRG which is a close relative of the DMRG.
Recall the infinite-system algorithm of the DMRG~\cite{orig-DMRG}.
 Write the ground-state wave function for $2N$-site
 system under the free boundary condition, obtained by diagonalizing
 the Hamiltonian matrix, as
\begin{equation}
\Psi_{G}^{2N}(\alpha|\beta) \label{eq:2}
\end{equation}
where $\alpha$ (resp. $\beta$) is the block-state index for the left-half
 (resp. right-half) $N$ sites.  For simplicity, we restrict ourselves to
 the cases where ${\cal H}$ is mirror symmetric and the ground-state
wavefunction
 is of even parity with respect to the space reflection.  If we extend
 the system
 by adding two sites at the center of the system, the ground-state wavefunction
 of the two-site extended system (with the extended Hamiltonian) can
 be written as
\begin{equation}
\Psi_{G}^{2N+2}(\alpha,i|j,\beta) \label{eq:3}
\end{equation}
where $i$ and $j$
 denote spin states (e.g., $i=-S,-S+1,\ldots,S$ in the
 $S^{z}$-diagonal representation) for added sites.  In the DMRG, we form the
 density matrix $\rho$ as
\begin{equation}
\rho(i',\alpha'|i,\alpha)
        = \sum_{j,\beta}
                \Psi_{G}^{2N+2}(\alpha',i'|j,\beta)
                \Psi_{G}^{2N+2}(\alpha, i |j,\beta)\label{eq:4}
\end{equation}
and diagonalize it to choose $m$ eigenstates with eigenvalues $\lambda_{\mu}$
 ($\mu=1,2,\ldots,m$, in descending order in magnitude).  We retain the
 $m$ eigenstates as new block states (bases) forming a truncated basis set
 for $(N+1)$-site system. These two, (P1) the diagonalization of ${\cal H}$ for
the extended
 system, and (P2) the choice of retained bases through the
density-matrix diagonalization,
 are the key processes in the DMRG.

In the PWFRG, these key processes are changed as follows. The process
 (P1) is replaced by, (P$1^{\prime}$) the {\em
improvement} of the ``input'' wavefunction. In 2D classical systems
 for which the original PWFRG is
 developed,
 the improvement is made by the multiplication of the transfer
 matrix~\cite{PWFRG}.
 For 1D quantum systems, the transfer-matrix multiplication is
successfully substituted
 by the modified Lanczos operation~\cite{mod-L1,mod-L2}
 on the wavefunction (see eq.(\ref{eq:9}) below).  As for
 the process (P2), the diagonalization of the density matrix is replaced
 by, (P$2^{\prime}$) diagonalization of
 the ``wavefunction matrix'' $\hat{\Psi}$:
\begin{equation}
\hat{\Psi}(\alpha,i|\beta,j)=\Psi_{G}(\alpha,i|j,\beta), \label{eq:5}
\end{equation}
where we have suppressed the superscript denoting the number of lattice
 sites.  In terms of the ``projection matrix'' $R(\alpha,i|\mu)$
 which projects
 the state $(\alpha,i)$
 into the new block state $\mu$, we have
\begin{equation}
\hat{\Psi}(\alpha,i|\beta,j)
        =\sum_{\mu}
                 R(\alpha,i|\mu)\, \omega_{\mu}\, R(\beta,j|\mu).
\label{eq:6}
\end{equation}
Since $\rho=\hat{\Psi}^2$, the density-matrix eigenvalues $\{\lambda_{\mu}\}$
 relate to the wavefunction-matrix eigenvalues $\{\omega_{\mu}\}$ as
$\lambda_{\mu}=\omega_{\mu}^2$~\cite{PWFRG}.
  Hence the choice of retained bases according to the magnitude
 of $|\omega_{\mu}|$
 is equivalent to that made in the DMRG.
As in the DMRG, the projection matrix
 $R$ is also used in the construction of the new block Hamiltonian
 for the size-extended system.

  There is one important process added in the PWFRG:
(P$3^{\prime}$) We set up a recursion
 relation for $R$ as follows. By $A(\alpha,i|\mu)$
 we denote the ``refined''
 projection matrix associated with the improved wavefunction obtained by
 the process (P$1^{\prime}$). The recursion relation reads~\cite{PWFRG}\\
\begin{equation}
 R_{{\rm new}}(\alpha,i|\mu)
        =\sum_{j,\eta,\xi}
               A(\xi,j|\alpha)
               R_{{\rm old}}(\xi,j|\eta)
               A(\eta,i|\mu).\label{eq:7}
\end{equation}
Updated wavefunction matrix is then given by
\begin{equation}
\hat{\Psi}_{{\rm new}}(\alpha,i|\beta,j)
        =\sum_{\mu}
                R_{{\rm new}}(\alpha,i|\mu)
                \, \omega_{\mu} \,
                R_{{\rm new}}(\beta,j|\mu), \label{eq:8}
\end{equation}
which we put again into the process (P$1^{\prime}$)
so that we can complete one iteration sequence.

 We summarize the algorithm of the PWFRG for 1D quantum
 system as follows:

\begin{itemize}
\item[Step 0.] Prepare an input wavefunction $\Psi_{{\rm old}}$ and an
associated
 initial projection matrix $R_{{\rm old}}$.
  Choice of the initial $\Psi_{{\rm old}}$
 is somewhat arbitrary; the exact ground-state wavefunction for a small-size
 system is conveniently chosen. Diagonalize the wavefunction matrix to obtain
 the initial projection matrix $\{R_{{\rm old}}(\mu|i,\alpha)\}$.
\newpage
\item[Step 1.] Apply the modified Lanczos operation $\hat{L}$ to $\Psi_{{\rm
 old}}$ to obtain the improved wave function $\Psi_{{\rm imp}}$. The operation
 $\hat{L}=\hat{L}(\alpha)$ on a state vector $|\Psi>$ is defined as
\begin{eqnarray}
\hat{L}(\alpha)|\Psi>&=&\frac{|\Psi>+\alpha|\chi>}
{\sqrt{1+\alpha^2<\chi|\chi>}}\nonumber \\
 &\equiv& |\tilde{\Psi}(\alpha)>\nonumber \\
|\chi>&\equiv&\Delta H|\Psi> \nonumber \\
\Delta H&\equiv&H-<\Psi|H|\Psi>.\label{eq:9}
\end{eqnarray}
The optimum choice $\alpha=\alpha^{*}$ minimizing
$<\tilde{\Psi}(\alpha)|H|\tilde{\Psi}(\alpha)>$ is given by
\begin{eqnarray}
\alpha^{*}
&=&\left(\chi_{3}-\sqrt{\chi_{3}^2+4\chi_{2}^3}\right)/(2\chi_{2}^2)
\nonumber \\
\chi_{n}&\equiv& <\Psi|(\Delta H)^{n}|\Psi>.\label{eq:10}
\end{eqnarray}
Write $\hat{L}(\alpha^{*})$ as $\hat{L}^{*}$.  In most cases we adopt $k$-fold
 operation with $k \ge2$:
\begin{eqnarray}
 |\Psi_{{\rm imp}}>
&=&\left[\hat{L}^{*}\right]^{k}|\Psi_{{\rm old}}>\nonumber\\
&=&\hat{L}^{*}(\hat{L}^{*}(\cdots(\hat{L}^{*}|\Psi_{{\rm
 old}}>))\cdots) \quad(\mbox{$k$ times}).\label{eq:11}
\end{eqnarray}
\item[Step 2.] Perform an eigenvalue decomposition of the wavefunction matrix
 $\{\hat{\Psi}_{{\rm imp}}(\alpha,i|\beta,j)\}$
 associated with the improved
 state vector $|\Psi_{{\rm imp}}>$:
\begin{equation}
\hat{\Psi}_{{\rm imp}}(\alpha,i|\beta,j)
        =\sum_{\mu}
                 A(\alpha,i|\mu)\, \omega_{\mu}\, A(\beta,j|\mu),\label{eq:12}
\end{equation}
where the new block index $\mu$ ($=1,2,\ldots$) is introduced in descending
 order of $|\omega_{\mu}|$ ($|\omega_{1}|\geq|\omega_{2}|\geq \cdots$). We
 then truncate the number of retained bases to $m$.
\item[Step 3.] Extend the system size by adding two sites at the
center of the chain.  Then using the ``refined'' projection matrix
$\{A(\alpha,i|\mu)\}$ ($\mu=1,2,\ldots,m$),
 update the left and right block Hamiltonian matrix just as in the DMRG.

\item[Step 4.] Update the projection matrix following eq.(\ref{eq:7})
 and form
 the new wavefunction $\Psi_{{\rm new}}$ according to eq.(\ref{eq:8}).
  Rename $\Psi_{{\rm
 new}}$ as $\Psi_{{\rm old}}$, then go to Step 1.
\end{itemize}
After repeating the above steps sufficiently many times ,
 we obtain the ``fixed point'' wavefunction.
  Physical quantities are calculated as the expectation
 value with respect to the fixed-point wavefunction.  In particular, the
 per-site magnetization $M$ is calculated as the expectation value of the
 center spins.

Let us give a comment on the relation between the PWFRG and the DMRG.  As
 regards the diagonalization of the wavefunction matrix,  a similar
 process (singular-value decomposition, SVD for short) appears in the
 original DMRG~\cite{orig-DMRG}.  The associated projection matrix $R$ is an
important object in both methods.  Main difference in the two methods
 lies in the updation process of $R$. In the DMRG, the updation
 $R_{\rm old}\rightarrow R_{\rm new}$ is ``directly'' given by the SVD
 of the updated ground-state wavefunction of the updated Hamiltonian
 matrix.  Whereas, in the PWFRG, the updation is given somewhat
 ``indirectly'' through the recursion relation eq.(\ref{eq:7}).
We {\em do not} determine $R_{\rm new}$ from the updated wavefunction, but we
{\em define}  $R_{\rm new}$  by eq.(\ref{eq:7}) and use it to
{\em form} the updated wavefunction.  The updated wavefunction is, then,
 a good guess of the ground-state wavefunction of the system with the
 extended size (by two sites).  We then refine the wavefunction by the
 modified Lanczos operation eq.(\ref{eq:9}) to obtain the ``refined
 projection matrix'' $A$.  As the iteration proceeds, the system size
 becomes larger and larger, and {\em at the same time}, the wavefunction
 becomes closer and closer to the ground-state wavefunction.
  The recursion relation eq.(\ref{eq:7}) implies that, at the fixed point
 (=infinite iterations), $R=A$ should be satisfied, and this is precisely
 the condition for the ground-state wavefunction which must be invariant
 under the modified Lanczos operation.

We apply the PWFRG to several models of quantum spin chains to draw the
 zero-temperature $M-H$ curves.  We present the results in the followings.

\begin{flushleft}
(a) $S=1/2$ XY antiferromagnetic chain~\cite{exact-XY1,exact-XY2}
\end{flushleft}

The local Hamiltonian $h$ (see eq.(\ref{eq:1})) for this model is given by
\begin{equation}
h(\vec{s}_i,\vec{s}_{j})=|J|(s_i^{x}s_{j}^{x}+s_i^{y}s_{j}^{y}),\label{eq:13}
\end{equation}
with ${\vec{s}_i}$
being $\frac{1}{2} \times $(Pauli spin matrices). This
 model is exactly soluble via the Jordan-Wigner transformation;
  the magnetization curve is known to be
\begin{equation}
M=\frac{1}{\pi} \sin^{-1}(\frac{H}{|J|}). \label{eq:14}
\end{equation}
This model is a gapless system even at $H=0$, which is reflected in the
 behavior of the $M-H$ curve : $\lim_{M\rightarrow
0+}H(M)=0$.  In
Fig.~\ref{fig:1}
 we compare the PWFRG calculation with the exact
solution, where an excellent
 agreement is seen. It should  be noted that even a small number
 ($m=16$) of retained bases gives the quite accurate $M-H$ curve.

\begin{flushleft}
(b) Takhtajan-Babujian model~\cite{TB1,TB2} ($S=1$)
\end{flushleft}

This model belongs to a family of models called bilinear-biquadratic chains
 whose local Hamiltonians have the form
\begin{equation}
h(\vec{s}_i,\vec{s}_j)
        =|J|[\vec{s}_i \cdot \vec{s}_j
         +\tilde{\beta}(\vec{s}_i\cdot\vec{s}_j)^2],\label{eq:21}
\end{equation}
where $\vec{s}_{i}$ and $\vec{s}_{j}$
are $S=1$ spin operators. The Takhtajan-Babujian
 model corresponds to $\tilde{\beta}=-1$, which is exactly soluble and is known
 to be gapless~\cite{TB1,TB2}.
  The $M-H$ curve can be drawn by solving the Bethe
 ansatz integral equation for the ``two-string'' (two-down-spin bound state)
 root density to obtain the ground-state energy density with fixed per-site
 magnetization $M$. For general $M$ the solution can only be obtained
numerically,
 by converting the integral equation into a matrix equation. As can be seen
 in Fig.~\ref{fig:3},
 our PWFRG calculation
reproduces the exact curve within a satisfactory
 precision even with the small number $m$ ($=20$) of the retained bases.

\begin{flushleft}
(c) Affleck-Kennedy-Lieb-Tasaki (AKLT) chain~\cite{AKLT} ($S=1$)
\end{flushleft}

This model also belongs to the family of the $S=1$ bilinear-biquadratic
 chains, whose local Hamiltonian being given by eq.(\ref{eq:21})
 with $\tilde{\beta}=1/3$. The
 ground state at $H=0$ is exactly known to be singlet and
gapped (the excitation
 gap $\Delta>0$), and is in the ``valence bond solid'' (VBS) state whose
 wavefunction is
 the product of matrices with finite
 dimensions~\cite{MP1,MP2}.
 The ground-state correlation length and the string order-parameter
 are also exactly known, from which the state is identified to be in the
 ``Haldane phase''~\cite{Haldane}.
 Increasing the magnetic field $H$, this
 VBS state remains to be the ground state up to the ``lower critical field''
 $H_{c1}$ above which the ground state is magnetized.  The critical field
 relates to the gap $\Delta$ as
\begin{equation}
H_{c1}=\Delta ,\label{eq:22}
\end{equation}
which holds for a general class of antiferromagnets so long as $\Delta$
 is the singlet-to-triplet energy gap.  There is another critical field
 $H_{c2}$ (upper critical field) above which the system's
per-site magnetization
 saturates to unity (complete ferromagnetic state).  For the
intermediate region
 of the field, $H_{c1}<H<H_{c2}$, the exact ground-state wavefunction has not
 been known even for the AKLT chain. It should be noted that the VBS-state
 wavefunction~\cite{AKLT,MP1,MP2}
 with finite matrix size $m$ ( $m=2$ for the AKLT chain ) is exactly
the fixed point of the iterations in the DMRG/PWFRG with the
number of retained bases
 just being $m$~\cite{OR}.
  Hence, by continuity, we can expect that
 the fixed-point wavefunction of the DMRG/PWFRG with relatively
 small $m$ should
 remains to be a good approximation to the correct ground-state wavefunction
 of the AKLT chain even in the intermediate range of the field.

In Fig.~\ref{fig:4} we draw the $M-H$ curve
 obtained by the PWFRG with
a small number $m$ ($=20$) of the retained bases.
The calculation
  reproduces the exact value of $H_{c2}=4|J|$ which can be derived from
 the stability consideration of the saturated state against the one down-spin
 formation~\cite{near-saturation1,near-saturation2,near-saturation3}.
 Further, the predicted square-root
behavior~\cite{near-saturation1,near-saturation2,near-saturation3}
\begin{equation}
M(H)\sim S-A_{2}\sqrt{H_{c2}-H}  \qquad ( ( 0 < ) A_{2}:{\rm constant} )
\label{eq:23}
\end{equation}
is
reproduced
(Fig.~\ref{fig:5}).

  As for the behavior near the lower critical field $H_{c1}$,
 another square-root
 behavior has been
expected~\cite{sqrt-behavior1,sqrt-behavior2,sqrt-behavior3,sqrt-behavior4}:
\begin{eqnarray}
M(H)&\sim& A_{1}\sqrt{H-H_{c1}}
        \qquad ( ( 0 < ) A_{1}:{\rm constant} ),\nonumber\\
H_{c1}&=&\Delta. \label{eq:24}
\end{eqnarray}
Although actual studies concerning this behavior have been almost limited
 to the case of the ``pure'' $S=1$ Heisenberg chain without the biquadratic
 term ($\tilde{\beta}=
0$ in eq.(\ref{eq:21})),
 the theoretical arguments for this behavior are
 so general that we can naturally expect the square-root behavior
 eq.(\ref{eq:24}) to
 hold for the AKLT
chain either.  As can be seen from Fig.~\ref{fig:6},
our PWFRG calculation is consistent with
 eq.(\ref{eq:24}); the calculated value
 of $H_{c1}\simeq0.7|J|$
 is in agreement with the previous numerical
 estimation of $\Delta$~\cite{AKLT-gap1,AKLT-gap2}.\\

To summarize,
in this article we have presented the application of the product-wavefunction
 renormalization-group (PWFRG) method~\cite{PWFRG}
 to quantum spin chains to draw the
 zero-temperature magnetization curve. The results show that the method
 works efficiently even for systems with finitely-magnetized ground states
 which are gapless.

We should give a comment on application of the DMRG to the magnetization
 process of quantum spin chains.  In a preliminary stage of the present
 study, we adopted the original infinite-algorithm of the DMRG.
  We then found that the DMRG, applied to quantum
 antiferromagnets in the magnetized region, is very likely to be ``trapped''
 by  ``metastable states'', leading typically to oscillatory behavior in
 the course of the iteration.  As a result, in most cases, we failed in
 obtaining the magnetized ground-state wavefunction corresponding to the
 fixed point.
In contrast,
 the PWFRG
 adopted in the present study is fairly stable, always successful in reaching
 the fixed point even in the magnetized region.
The above ``instability'' of the DMRG  is certainly related to the nature of
 the present problem.  The magnetization curve reflects a continuous sequence
 of level-crossing transitions induced by varying magnetic field, between
 the states with different values of total $S^{z}$.  The exact diagonalization
 (instead of the modified Lanczos operation used in the present study)
 in the original DMRG extracts the exact lowest energy state
of the truncated Hamiltonian
at each iteration step.  The state is, however, the lowest energy state of
 the system at that size ($\sim2\times\mbox{iteration steps}$); owing to the
 highly degenerate energy level structure, there may well occur ``level
 crossings'' on the increasing system size,
 leading to the oscillatory behavior
 in the iteration.

 Of course, we can remove this instability
 by decomposing the state space
 into subspaces according to quantum numbers to reduce the degeneracy. For
 the problem of the magnetization process the quantum number is the total
$S^{z}( \equiv S^{z}_{\rm T} )$ and we can surely perform the DMRG
 with fixed $S^{z}_{\rm T}$.  This approach, however,
 is rather laborious because what we need is the ground-state energy density
 with fixed magnetization {\em density} $S^{z}_{\rm T}/N$
 ($N$: lattice size), which
 requires a huge number of fixed-$S^{z}_{\rm T}$ calculations
 ($S^{z}_{\rm T}=0,1,2,\ldots,NS$, for the spin $S$ case)
 for large $N$, to find the optimum value of $S^{z}_{\rm T}$
 giving the true ground state under the field.
 On the other hand, with the PWFRG method,
 the system gradually and automatically
 converges into the ``true'' ground state under the given magnetic field.
 Due to this ``stability'', the PWFRG will be particularly
 useful for ``critical'' systems whose energy spectrum
(or transfer-matrix eigenvalue spectrum, for 2D classical systems)
is highly degenerate.

The modified Lanczos method used in this article is a two-step restriction
of the ordinary Lanczos algorithm.  Similar restricted-step Lanczos has
 been also employed in a new version of the DMRG
 (finite-system algorithm)~\cite{new-DMRG} which shows very high
 efficiency. Although precise relation between the PWFRG and the new DMRG is
not obvious, we can say that the former may, in a sense, be the
 infinite-system-algorithm version of the latter.

As for the magnetization process, the ``middle-field'' phase
transitions to occur between $H_{c1}$ and $H_{c2}$ have been known
for some systems~\cite{Parkinson,Kiwata-Akutsu,Sato-Akutsu}.
Study of such field-induced ground-state phase transitions is an
interesting future problem.  Finite-temperature
behavior~\cite{Yamamoto-Miyashita,B-X-G} of the magnetization process
is also an interesting problem.

We would like to thank T. Nishino, M. Kikuchi, H. Kiwata and R. Sato
 for valuable discussions.
This work was partially supported by the Grant-in-Aid for Scientific
Research from Ministry of Education, Science, Sports and Culture
(No.07640514).  A part of the numerical computation was made on the
system VPP500 of the Supercomputer Center, Institute for Solid State
Physics, the University of Tokyo. One of us (K.O) is supported by JSPS
Research Fellowships for Young Scientists.

%
%
%
\newpage
\begin{figure}
        \vspace*{1cm}
        \caption{$M-H$ curve of the $S=1/2$ XY antiferromagnetic chain.}
        \label{fig:1}
\end{figure}
\begin{figure}
        \vspace*{1cm}
        \caption{$M-H$ curve of the Takhtajan-Babujian model.}
        \label{fig:3}
\end{figure}
\begin{figure}
        \vspace*{1cm}
        \caption{$M-H$ curve of the Affleck-Kennedy-Lieb-Tasaki chain.}
        \label{fig:4}
\end{figure}
\begin{figure}
        \vspace*{1cm}
        \caption{
                 The magnetization curve
                 of the Affleck-Kennedy-Lieb-Tasaki chain
                near the saturation field $H_{c2}$.}
        \label{fig:5}
\end{figure}
\begin{figure}
        \vspace*{1cm}
        \caption{
                 The magnetization curve
                 of the Affleck-Kennedy-Lieb-Tasaki chain
                near the lower critical field $H_{c1}$.}
        \label{fig:6}
\end{figure}

\end{document}